\documentclass[12pt]{article}
\title{On Modal Logics of Partial Recursive Functions}
\author{Pavel Naumov\\ \\
	Computer Science\\
	Pennsylvania State University\\
	Middletown, PA 17057\\ \\
	{\sf naumov@psu.edu}}
\usepackage{amssymb}
\usepackage{amsmath}
\usepackage{graphics}
\usepackage{natbib}
\newtheorem{definition}{Definition}
\newtheorem{theorem}{Theorem}
\newtheorem{lemma}{Lemma}

\newtheorem{corollary}{Corollary}
\newenvironment{proof}{\noindent{\sf Proof.}}{\hfill $\Box$\linebreak}
\newsavebox{\doublewedgesavebox}
\savebox{\doublewedgesavebox}[3.5mm]{$\wedge\hspace{-1.5mm}\wedge$}
\newcommand{\doublewedge}{\usebox{\doublewedgesavebox}}

\begin{document}

\maketitle

\begin{abstract}
	The classical propositional logic is known to be sound and complete with
	respect to the set semantics that interprets connectives as
	set operations. The paper extends propositional language by a new binary
	modality that corresponds to partial recursive function type constructor
	under the above interpretation. The cases of deterministic and 
	non-deterministic functions are considered and for both of them
	 semantically complete modal logics are described and decidability of
	 these logics is established.

	 \vspace{1mm}
\end{abstract}

\section{Introduction}

We are interested in the use of logical connectives to describe properties of the set and type operations. Historically, there have been two major ways to interpret logical connectives as such operations: Curry-Howard isomorphism and set semantics. 

Under Curry-Howard isomorphism (\cite{Curry1934,MR3:289d,MR20:817,MR82g:03094}), propositional formulas are interpreted as types and connectives $\wedge, \vee$, and $\rightarrow$ are interpreted as Cartesian product, disjoint union, and constructive function type constructors. It can be shown that a formula is provable in intuitionistic propositional logic if and only if it is always evaluated to an inhabited type. Thus, intuitionistic logic could be viewed as a calculus that describes properties of Cartesian product, disjoint union, and function type constructors. 

Since the list of possible type constructors is not limited to just the trio of product, disjoint union, and function, one can raise a question about logical principles describing behavior of other type constructors. For example, list, partial object \citep{Smith1995} and squash \citep{Constable1986} types can be viewed as modalities while inductive and co-inductive constructors (\cite{MR92h:03020} and \cite{MR91h:03012}) may be considered as quasi-quantifiers. In fact, \cite{Kopylov2001} established that modal logic of squash operator is Lax Logic \citep{MR98j:03045}.

According to the set semantics, every propositional formula is evaluated to a subset of a given universe $U$ and propositional connectives conjunction $\wedge$, disjunction $\vee$, and negation $\neg$ are identified with set operations intersection $\cap$, union $\cup$, and complement $\complement_U$, correspondingly. It is easy to see that a formula is provable in the classical propositional logic if and only if it is evaluated to the entire universe $U$ under any interpretation of propositional variables.

Several possible extensions of the classical logic by modal operators corresponding, under the above set semantics, to additional set operations have been considered. \cite{MR5:211f} established that if the universe $U$ is a topological space, then modal logic S4 describes properties of the interior operator. If the universe $U$ is the set of all words in some alphabet, then properties of the logical connectives corresponding to product and star operations are axiomatized by Interval Temporal Logic \citep{MR778950}. In \citep{Naumov03, Naumov04}, the author describes an extension of the classical propositional logic by binary modalities, corresponding to the operations disjoint union and Cartesian product.

This paper considers an extension of the classical propositional logic by a binary modality $\rhd$, corresponding to computable function type constructor. Namely, if $U$ is the universe of all words in some alphabet, then $(\phi\rhd\psi)^*$ is the set of all Turing machine descriptions of partial recursive functions from $\phi^*$ into $\psi^*$. We consider cases of deterministic and nondeterministic Turing machines. For both of them complete Hilbert-style axiomatizations of the appropriate modal logics is given. It turns out that modal logic of deterministic functions $\Re_d$ is an extension of the modal logic of nondeterministic functions $\Re$ by just one additional axiom. 

The modality $\phi\rhd\psi$ of the logics of partial recursive functions is, essentially,  a form of Hoare triple $\phi\{\alpha\}\psi$ with a fixed program variable $\alpha$. Thus, there is some similarity between modal logics of recursive functions and the dynamic logic \citep{MR2001i:68110}. For example, introduced below axiom of logic $\Re$: 
$\phi\rhd\psi\rightarrow(\chi\rhd\psi\rightarrow(\phi\vee\chi)\rhd\psi)$ could be related to dynamic logic theorem $\phi\{\alpha\}\psi \rightarrow(\chi\{\alpha\}\psi\rightarrow(\phi\vee\chi)\{\alpha\}\psi)$.
This similarity, however, ends once iterative applications of the modality are considered. For example, formula $(\top\{\alpha\}\phi)\{\alpha\}\phi$ is also a theorem of the dynamic logic but modal formula $(\top\rhd\phi)\rhd\phi$ is valid neither in $\Re$ nor in $\Re_d$.

This paper focuses on soundness and completeness of logics $\Re$ and $\Re_d$ with respect to the class of partial recursive functions. As one can expect, the results can be easily relativized by an oracle. It is worth mentioning, although, that presented in the paper soundness and completeness proofs could also be adopted for some subclasses of the class of partial recursive functions such as, for example, polynomial functions and finite-domain functions. Hence, both of these logics capture very general properties of ``complete", in some informal sense, classes of enumerable functions. The downside of this, of course, is that more specific properties of recursive functions are not reflected in these logics. For example, many of the properties of recursive functions captured by the intuitionistic logic under Curry-Howard isomorphism, such as closure under composition, could not be expressed in logics $\Re$ and $\Re_d$. One should think of these logics more as an attempt to reason about functions in (a modal extension of) the classical propositional logic rather than a modal axiomatization of recursiveness. Similarly defined logics of {\em total} recursive functions, as will be mentioned in the conclusion, would provide a significantly more expressive language. Our investigation of logics $\Re$ and $\Re_d$ could be viewed as a first step towards study of such more expressive logics.

The results for logics $\Re$ and $\Re_d$ will be presented together. In the next section we discuss the definition of the recursive functions and the Kleene recursion theorem on which our completeness results are based. In Section 3, a formal semantics of the modal logics of recursive functions is given. Section 4 lists axioms and inference rules for both logics and verifies their soundness. The rest of the paper is dedicated to the completeness proof. In Section 5, Kripke-style models for $\Re$ and $\Re_d$ are introduced and completeness of these logics with respect to appropriate classes of the Kripke models is proven. In Section 6, in order to finish the proof of the completeness theorem, we show how Kripke models could be converted into sets of partial recursive functions. Decidability of the logics follows from finiteness of the corresponding Kripke models. Section 7 concludes with the discussion of an alternative definition of the logic of nondeterministic partial recursive functions and the logics of total recursive functions.

\section{Recursive functions}

We study modal logic descriptions of partial recursive functions. The two classes of recursive functions -- deterministic and nondeterministic -- will be considered. Nondeterministic partial recursive functions could be described, for example, as nondeterministic Turing machines. Value $f(x)$ of a nondeterministic function $f$ on an argument $x$ is defined as the set of all values that a nondeterministic machine representing $f$ can return on input $x$.
Deterministic partial recursive function is a special case of nondeterministic function whose value is a set that has no more than one element. 

We consider an enumeration $\{\xi_u\}_{u\in U}$ of partial recursive functions from a universe $U$ into $U$ by the elements of the same universe $U$. The two major cases that will be considered are: a) $\{\xi_u\}_{u\in U}$ is an enumeration of all nondeterministic partial recursive functions and b) $\{\xi_u\}_{u\in U}$ is an enumeration of all deterministic partial recursive functions. The exact choice of the universe and the enumeration will not be important as long as the following version of Kleene's recursion theorem is satisfied:

\begin{theorem}\label{Kleene}
For any finite set $f_1,\dots,f_n$ of total recursive functions from $U^n$ to $U$ there are elements $u_1,\dots,u_n\in U$ such that $\xi_{u_i} \equiv \xi_{f_i(u_1,\dots,u_n)}$ for  any $0\le i\le n$.
\end{theorem}
Note that reproduced below standard (see, for example, \cite{MR88b:03059}) proof of the recursion theorem for enumeration $\{\xi_u\}_{u\in U}$ of deterministic partial recursive functions is also valid for enumerations of nondeterministic partial recursive functions.

\begin{proof}
Let $\{\delta^n_x\}_{x\in U}$ be an enumeration of deterministic partial recursive functions of arity $n$ by elements of the universe $U$. 
Consider recursive functions $g_i:U^n\mapsto U$ such that for any $x_1,\dots,x_n\in U$,
$$\xi_{g_i(x_1,\dots,x_n)}(y) = \left\{ \begin{array}{ll}
		\xi_{\delta^n_{x_i}(x_1,\dots,x_k)}(y) & 
			\mbox{if  $\delta^n_{x_i}(x_1,\dots,x_k)$ convergent}\\
		\mbox {divergent} & \mbox{otherwise}
		\end{array} \right.$$
Note that $h_i(x_1,\dots,x_n) = f_i(g_1(x_1,\dots,x_n),\dots,g_n(x_1,\dots,x_n))$ is a total recursive function $U^n\mapsto U$ for any $i$. Let $w_i$ be such that $\delta^n_{w_i}\equiv h_i$. Thus,
$$\xi_{g_i(w_1,\dots,w_n)}\equiv
\xi_{\delta^n_{w_i}(w_1,\dots,w_n)}\equiv
\xi_{h_i(w_1,\dots,w_n)} \equiv
\xi_{f_i(g_1(w_1,\dots,w_n),\dots,g_n(w_1,\dots,w_n))}
.$$
Take $u_i$ to be $g_i(w_1,\dots,w_n)$.
\end{proof}

\section{Semantics}

\begin{definition}\label{language}
The formulas of the modal language $\cal L$ are built from propositional variables $p, q, r \dots$ and false constant $\bot$ using implication $\rightarrow$ and binary modality $\rhd$.
\end{definition}
As usual, boolean connectives conjunction $\wedge$, disjunction $\vee$,  negation $\neg$, and constant true $\top$ are assumed to be defined through implication and false.  Let $\doublewedge\Gamma$ be the conjunction of all formulas from a finite set $\Gamma$. By definition, $\doublewedge\varnothing$ is $\top$.

\begin{definition}\label{semantics}
Valuation $*$ is an arbitrary mapping of propositional variables into subsets of the universe. We define mapping $(\cdot)^*$  that extends $*$ to a mapping from modal propositional formulas into subsets of $U$:
\begin{enumerate}
\item 	$\bot^* = \varnothing$,
\item 	$(\phi\rightarrow\psi)^* =
	 \complement_U(\phi^*)\cup\psi^*$,
\item $(\phi\rhd\psi)^* = \{w\in U \;|\; \forall u\in \phi^*\;(\xi_w(u)\neq\varnothing \rightarrow \xi_w(u)\cap\psi^*\neq\varnothing)\}$.
\end{enumerate} 
If $\phi^*=U$ for any valuation $*$, then we say that propositional modal formula $\phi$ is a tautology of enumeration $\{\xi_u\}_{u\in U}$. Notation: $\{\xi_u\}_{u\in U}\vDash\phi$. 
\end{definition}
Part three of the above definition stipulates that a nondeterministic function belongs to $(\phi\rhd\psi)^*$ if for any argument from $\phi^*$, on which this function is defined, at least one of its values belongs to $\psi^*$. An alternative definition, when all such values are required to belong to $\psi^*$, is discussed in the conclusion.

\section{Axioms}

\begin{definition}
The modal logic $\Re$ of nondeterministic partial recursive functions is an extension of the classical propositional logic, formulated in the language $\cal L$, by the following axioms 
\begin{itemize}
\item[A1.] 	
$\phi\rhd\psi\rightarrow(\chi\rhd\psi\rightarrow(\phi\vee\chi)\rhd\psi)$,
\item[A2.] $\bot\rhd\phi$,
\item[A3.] $\phi\rhd\top$,
\end{itemize}
and, in addition to Modus Ponens, the following monotonicity inference rule:
\begin{itemize}
\item[M.]$\dfrac{\phi_1\rightarrow\phi_2,\;\;\;\;\;\psi_1\rightarrow\psi_2}
{\phi_2\rhd\psi_1\rightarrow \phi_1\rhd\psi_2}$
\end{itemize}
\end{definition}

\begin{definition}
The modal logic $\Re_d$ of deterministic partial recursive functions, in addition to the axioms and the inference rules of $\Re$, contains the following additional axiom:
\begin{itemize}
\item[A4.] $\phi\rhd\psi\rightarrow(\phi\rhd\chi\rightarrow\phi\rhd(\psi\wedge\chi))$.
\end{itemize}
\end{definition}
Let $\Delta \vdash_L \phi$ mean that formula $\phi$ is provable from a set of formulas $\Delta$ and the theorems of modal logic $L$ using only Modes Ponens inference rule.

\begin{lemma}\label{xor_left}
$$(a\wedge c)\rhd b, (a\wedge\neg c)\rhd b \vdash_\Re a\rhd b$$
\end{lemma}
\begin{proof}
Assume $(a\wedge c)\rhd b$ and $(a\wedge \neg c)\rhd b$. By axiom A1, 
\begin{equation}\label{xor_left_eq}
((a\wedge c)\vee(a\wedge\neg c))\rhd b.
\end{equation}
On the other hand, since $a \rightarrow (a\wedge c)\vee(a\wedge\neg c)$ is a propositional tautology, by rule M, 
$$\vdash_\Re ((a\wedge c)\vee(a\wedge\neg c))\rhd b \rightarrow a\rhd b.$$
This, in combination with formula (\ref{xor_left_eq}), implies $a\rhd b$.
\end{proof}

\begin{lemma}\label{xor_right}
$$a\rhd\neg(b\wedge c), a\rhd\neg(b\wedge\neg c) \vdash_{\Re_d}a\rhd\neg b$$
\end{lemma}
\begin{proof}
Assume $a\rhd\neg(b\wedge c)$ and $a\rhd\neg(b\wedge\neg c)$. By axiom A4,
\begin{equation}\label{xor_right_eq}
a\rhd (\neg(b\wedge c) \wedge \neg(b\wedge\neg c))
\end{equation}
On the other hand, since $\neg(b\wedge c) \wedge \neg(b\wedge\neg c) \rightarrow \neg b$ is a propositional tautology, by rule M, 
$$\vdash_{\Re_d} a\rhd(\neg(b\wedge c) \wedge \neg(b\wedge\neg c)) \rightarrow a\rhd\neg b.$$
This, in combination with formula (\ref{xor_right_eq}), implies $a\rhd\neg b$.
\end{proof}

\begin{theorem}\label{soundness}
For any propositional modal formula $\phi$, 
\begin{enumerate}
\item If $\vdash_\Re\phi$, then $\{\xi_u\}_{u\in U}\vDash\phi$ for any enumeration $\{\xi_u\}_{u\in U}$ of nondeterministic recursive functions,
\item If $\vdash_{\Re_d}\phi$, then $\{\xi_u\}_{u\in U}\vDash\phi$ for any enumeration $\{\xi_u\}_{u\in U}$ of deterministic recursive functions.
\end{enumerate}
\end{theorem}
\begin{proof} Both parts of the theorem will be proven simulteniously by the induction on the size of the derivation of formula $\phi$. Cases of classical logic axioms and Modes Ponens inference rule are trivial. Let us consider axioms A1-A4 and the monotonicity rule M:
\begin{enumerate}
\item[A1.] Suppose that $w\in(\phi\rhd\psi)^*$ and $w\in(\chi\rhd\psi)^*$. We will show that $w\in((\phi\vee\chi)\rhd\psi)^*$. Indeed, assume that there is $u\in(\phi\vee\chi)^*$ such that $\xi_w(u)\neq\varnothing$. Note that $(\phi\vee\chi)^*=\phi^*\cup\chi^*$. Thus, $u\in\phi^*$ or $u\in\chi^*$. In the first case, because $w\in(\phi\rhd\psi)^*$, we can conclude that $\xi_w(u)\cap\psi^*\neq\varnothing$. Therefore, $w\in((\phi\vee\chi)\rhd\psi)^*$. The second case is similar.
\item[A2.] For any $w\in U$ and any valuation $*$, statement 
$$\forall u\in\bot^*\;(\xi_w(u)\neq\varnothing\rightarrow\xi_w(u)\cap\psi^*\neq\varnothing)$$
is true because $\bot^*=\varnothing$.
\item[A3.] For any $w\in U$ and any valuation $*$, statement 
$$\forall u\in\phi^*\;(\xi_w(u)\neq\varnothing\rightarrow\xi_w(u)\cap\top^*\neq\varnothing)$$
is true because $\top^*=U$.
\item[A4.] Applicable only to the second part of the theorem. Suppose that $w\in(\phi\rhd\psi)^*$ and $w\in(\psi\rhd\chi)^*$. We will show that $w\in(\phi\rhd(\psi\wedge\chi))^*$. Indeed, assume that there is $u\in\phi^*$ such that $\xi_w(u)\neq\varnothing$. Note that  $w\in(\phi\rhd\psi)^*$ and $w\in(\psi\rhd\chi)^*$ imply that $\xi_w(u)\cap\psi^*\neq\varnothing$ and $\xi_w(u)\cap\chi^*\neq\varnothing$. Since $\xi_w(u)$ cannot contain more than one element, $\xi_w(u)\cap(\psi^*\cap\chi^*)\neq\varnothing$. Therefore, $w\in(\phi\rhd(\psi\wedge\chi))^*$.
\item[M.] If $\phi_1^*\subseteq\phi_2^*$ and $\psi_1^*\subseteq\psi_2^*$, then any function from $\phi_2^*$ into $\psi_1^*$ is also a function from $\phi_1^*$ into $\psi_2^*$.
\end{enumerate}
\end{proof}

\section{Kripke Models}

\begin{definition}
Kripke model is a triple $\langle W, \rightarrow, \Vdash \rangle$, where $W$ is a finite set of ``worlds", $\rightarrow$ is a ternary ``computability" relation on worlds, and $\Vdash$ is a binary ``forcing" relation between worlds and propositional formulas.
\end{definition}
Informally, worlds should be viewed as program codes and $u\rightarrow_wv$ as a statement that program $w$ on input $v$ might terminate with output $v$.

\begin{definition}
Kripke model is called deterministic if for any worlds $w,u\in W$ there is no more than one $v\in W$ such that $u\rightarrow_wv$.
\end{definition}

\begin{definition}\label{forcing}
For any Kripke model the forcing relation is extended to relations $\Vdash$ between worlds and modal formulas as follows:
\begin{enumerate}
\item $w\nVdash\bot$,
\item $w\Vdash \phi\rightarrow\psi$ if and only if either $w\nVdash\phi$ or $w\Vdash\psi$,
\item $w\Vdash\phi\rhd\psi$ iff for any worlds $u$ and $v$ such that $u\rightarrow_wv$ and $u\Vdash\phi$ there is world $v'$ such that $u\rightarrow_wv'$ and $v'\Vdash\psi$.
\end{enumerate} 
\end{definition}
Note that in the case of a deterministic Kripke model, worlds $v$ and $v'$ in the above definition are the same.

\begin{theorem}\label{Kripke completeness}
For any propositional modal formula $\phi_0$,
\begin{enumerate}
\item  If $\nvdash_\Re\phi_0$, then there is a world $w$ of a Kripke model  $\langle W, \rightarrow, \Vdash \rangle$ such that $w\nVdash\phi_0$.
\item If $\nvdash_{\Re_d}\phi_0$, then there is a world $w$ of a deterministic Kripke model $\langle W, \rightarrow, \Vdash \rangle$ such that $w\nVdash\phi_0$.
\end{enumerate}
\end{theorem}
\begin{proof} Justifications of the two parts of this theorem are similar. We will present them in one proof. Let symbol $\vdash$ below stand for $\vdash_\Re$ or $\vdash_{\Re_d}$, depending on whether we prove the first or the second part of the theorem.
\begin{definition}
Let us define operation $\sim$ on modal propositional formulas as follows: $\sim(\neg\phi)$ is $\phi$ for any propositional modal formula $\phi$ and $\sim\phi$ is $\neg\phi$ if $\phi$ is not, syntactically, a negation of some formula.
\end{definition}
One can easily see that $\sim\phi$ is equivalent to $\neg\phi$ in the classical propositional logic. Since logics $\Re$ and $\Re_d$ are extensions of the classical logic, the same equality holds there too.  

\begin{definition}
Let $\Phi_0$ be a finite extension of $\{\phi_0\}$ closed with respect to subformulas and operation $\sim$.
\end{definition}

\begin{definition}
	For any subsets $u$, $v$, and $w$ of $\Phi_0$, pair $(u,v)$ is 
	$w$-consistent if 
	$w \nvdash \doublewedge u \rhd \neg \doublewedge v$.
\end{definition}

\begin{lemma}\label{consistency lemma}
	If pair $(u,v)$ is $w$-consistent, then sets $u$ and $v$ are consistent.
\end{lemma}
\begin{proof} Assume that $u$ is not consistent: $\vdash \doublewedge u \rightarrow \bot$. Thus, by rule M, we have
$\vdash \bot\rhd\neg \doublewedge v \rightarrow \doublewedge u \rhd \neg \doublewedge v$. 
Hence, by axiom A2, $\vdash \doublewedge u \rhd \neg \doublewedge v$. This contradicts to $w$-consistency of pair $(u,v)$. 

Next, suppose that $v$ is inconsistent: $\vdash\top \rightarrow \neg \doublewedge v$. Thus, by rule M, one can conclude that
$\vdash \doublewedge u \rhd \top \rightarrow \doublewedge u \rhd \neg \doublewedge v$. Taking into account axiom A3, $\vdash\doublewedge u \rhd \neg \doublewedge v$. Again contradiction with $w$-consistency of pair $(u,v)$. 
\end{proof}

\begin{lemma}\label{left extension lemma} 
For any $w$-consistent  pair $(u,v)$ of subsets of $\Phi_0$, subset $u$ can be extended to a complete  consistent subset $u'$ of $\Phi_0$ such that pair $(u',v)$ is still $w$-consistent.
\end{lemma}
\begin{proof} We only need to prove that for any formula $\phi$ either $\phi$ or $\neg\phi$ could be added to $u$ to keep pair $(u,v)$ consistent. Assume that $w \vdash (\doublewedge u\wedge\phi)\rhd\neg\doublewedge v$ and 
$w \vdash (\doublewedge u\wedge\neg\phi)\rhd\neg\doublewedge v$.
By Lemma \ref{xor_left}, 
$w \vdash \doublewedge u\rhd\neg\doublewedge v$.
Therefore, $(u,v)$ is not $w$-consistent. Contradiction. 
\end{proof}

\begin{lemma}\label{right extension lemma} 
For any $w$-consistent  in logic $\Re_d$ pair $(u,v)$ of subsets of $\Phi_0$, subset $v$ can be extended to a complete and consistent in $\Re_d$ subset $v'$ of $\Phi_0$ such that pair $(u,v')$ is still $w$-consistent in logic $\Re_d$.
\end{lemma}
\begin{proof}
Similarly to the proof of Lemma~\ref{left extension lemma}, assume that 
$w\vdash_{\Re_d} \doublewedge u\rhd\neg(\doublewedge v\wedge\phi)$ and
$w\vdash_{\Re_d} \doublewedge u\rhd\neg(\doublewedge v\wedge\neg\phi)$.
By Lemma \ref{xor_right}, 
$w\vdash_{\Re_d} \doublewedge u\rhd\neg\doublewedge v$.
Therefore, $(u,v)$ is not $w$-consistent. \end{proof}

\begin{definition}\label{canonical nu Kripke model} 
Let Kripke model $K=\langle W,\rightarrow,\Vdash\rangle$ be defined as follows: $W$ is the set of all pairs $(w,\phi)$ where $w$ is a maximal consistent in $\Re$ subset of $\Phi_0$ and $\phi$ is a formula from $\Phi_0$, $(u,\psi)\rightarrow_{(w,\phi)}(v,\chi)$ is true if $(u,\{\psi\})$ is a $w$-consistent in $\Re$ pair and $\psi\in v$, and $(w,\phi)\Vdash p$ is true if $p\in w$. 
\end{definition}

\begin{lemma}\label{main nu Kripke}
For any formula $\phi\in\Phi_0$ and any world $(w,\psi)$ of model $K$,
$$\phi \in w\;\;\;\;\;\Longleftrightarrow \;\;\;\;\; (w,\phi)\Vdash\phi.$$
\end{lemma}
\begin{proof}
Induction on the complexity of formula $\phi$. The only non-trivial case is when $\phi\equiv\phi_1\rhd\phi_2$. 
\begin{itemize}
\item[$\Rightarrow$] Suppose that $\phi_1\rhd\phi_2\in w$. Consider any world $(u,\psi)$ such that $(u,\psi)\Vdash\phi_1$. Case 1: $(u,\psi)$ is not $w$-consistent. Thus, by Definition~\ref{canonical nu Kripke model}, there is no $(v,\chi)$ such that $(u,\psi)\rightarrow_{(w,\phi)}(v,\chi)$. Therefore, $w\Vdash\phi_1\rhd\phi_2$. Case 2: $(u,\psi)$ is $w$-consistent. By the induction hypothesis, $\phi_1\in u$. Thus, $\vdash_\Re\doublewedge u\rightarrow\phi_1$. We will show that set $\{\phi_2,\psi\}$ is consistent. Indeed, if $\phi_2\vdash_\Re\neg\psi$, then, by rule M, we have $\vdash_\Re\phi_1\rhd\phi_2\rightarrow\doublewedge u\rhd\neg\psi$. Hence, $w\vdash_\Re\doublewedge u\rhd\neg\psi$. This means that pair $(u,\psi)$ is not $w$-consistent. Contradiction. Thus, $\{\phi_2,\psi\}$ is a consistent set. Let $v$ be its any consistent extension and $\chi$ be any formula of $\Phi_0$. By the induction hypothesis, $(v,\chi)\Vdash\phi_2$.
By Definition~\ref{canonical nu Kripke model}, 
$(u,\psi)\rightarrow_{(w,\phi)}(u,\chi)$.

\item[$\Leftarrow$]
Suppose that $\phi_1\rhd\phi_2\notin w$. By rule M, $w\nvdash_\Re \phi_1\rhd\neg\neg\phi_2$. Thus, pair $(\{\phi_1\},\{\neg\phi_2\})$ is $w$-consistent. By Lemma~\ref{left extension lemma}, there is a complete consistent extension $u$ of $\{\phi_1\}$ such that $(u,\{\neg\phi_2\})$ is $w$-consistent. By the induction hypothesis, $(u,\neg\phi_2)\Vdash\phi_1$. By Lemma~\ref{consistency lemma}, set $\{\neg\phi_2\}$ is consistent. Consider an arbitrary complete and consistent extension $v$ of this set and an arbitrary formula $\chi$ of $\Phi_0$. Trivially, $(u,\neg\phi_2)\rightarrow_{(w,\phi)}(u,\chi)$. At the same time, for any $(v',\chi')$ such that $(u,\neg\phi_2)\rightarrow_{(w,\phi)}(v',\chi)$ we will have $\neg\phi_2\in v'$ and, thus, by the induction hypothesis, $(v',\chi')\nVdash\phi_2$. Therefore, $(w,\phi)\nVdash\phi$.
\end{itemize}
\end{proof}

\begin{definition}\label{canonical Kripke model} 
Let deterministic Kripke model $K_d=\langle W,\rightarrow,\Vdash\rangle$  be defined as follows: $W$ is the set of all pairs of maximal consistent in $\Re_d$ subset of $\Phi_0$, $(u_1,u_2)\rightarrow_{(w_1,w_2)}(v_1,v_2)$ is true if $(u_1,v_1)$ is a $w_1$-consistent in $\Re_d$ pair and $u_2=v_1=v_2$, and $(w_1,w_2)\Vdash p$ is true if $p\in w_1$. 
\end{definition}

\begin{lemma}\label{main Kripke}
For any formula $\phi\in\Phi_0$ and any world $(w_1,w_2)$ of model $K_d$,
$$\phi \in w_1\;\;\;\;\;\Longleftrightarrow \;\;\;\;\; (w_1,w_2)\Vdash\phi.$$
\end{lemma}
\begin{proof} Induction on complexity of formula $\phi$. The only non-trivial case is when $\phi$ is $\phi_1\rhd\phi_2$ for some modal formulas $\phi_1$ and $\phi_2$. 
\begin{itemize}
\item[$\Rightarrow$] Assume that $\phi_1\rhd\phi_2\in w_1$. Consider an arbitrary $w_1$-consistent pair $(u,v)$ of maximal consistent subsets of $\Phi_0$. It will be sufficient to show that if $(u,v)\Vdash\phi_1$, then $(v,v)\Vdash\phi_2$. Indeed, assume that $(u,v)\Vdash\phi_1$ and  $(v,v)\nVdash\phi_2$. By the induction hypothesis, $\phi_1\in u$ and $\phi_2\notin v$. Thus, by maximality of $v$, we have $\sim\phi_2\in v$. Hence formulas $\doublewedge u\rightarrow\phi_1$ and $\phi_2\rightarrow\neg\doublewedge v$ are provable in the classical propositional logic. By rule M,
$\vdash_{\Re_d}\phi_1\rhd\phi_2 \rightarrow 
\doublewedge u\rhd\neg\doublewedge v$. Given that $\phi_1\rhd\phi_2\in w_1$, we can conclude that $w_1 \vdash_{\Re_d} \doublewedge u\rhd\neg\doublewedge v$. Therefore, $(u,v)$ is not a $w_1$-consistent pair. Contradiction. 

\item[$\Leftarrow$] Suppose $\phi_1\rhd\phi_2\notin w_1$. By maximality of $w_1$, we have  $w_1 \nvdash_{\Re_d} \phi_1\rhd\phi_2$. Thus, $(\{\phi_1\},\{\neg\phi_2\})$ is a $w_1$-consistent pair of sets. By Lemma \ref{left extension lemma} and Lemma \ref{right extension lemma}, it can be extended to a pair $(u,v)$ of maximal consistent sets which is also $w_1$-consistent. By Definition~\ref{canonical Kripke model}, $(u,v)\rightarrow_{(w_1,w_2)}(v,v)$. By the induction hypothesis, $(u,v)\Vdash\phi_1$ and $(v,v)\nVdash\phi_2$. Therefore, $(w_1,w_2)\nVdash\phi_1\rhd\phi_2$.
\end{itemize}
\end{proof} 
Let us finish the proof of the completeness theorem. If $\nvdash_\Re\phi_0$, then consistent subset $\{\sim\phi_0\}$ of $\Phi_0$ could be extended to a maximal consistent subset $w$ of $\Phi_0$. By Lemma~\ref{main nu Kripke}, $(w,\phi_0)\nVdash\phi_0$. Similarly, if $\nvdash_{\Re_d}\phi_0$, then $\{\sim\phi_0\}$ is consistent subset of $\Phi_0$. It can be extended to a maximal consistent subset $w$ of $\Phi$. By Lemma~\ref{main Kripke}, $(w,w)\nVdash\phi_0$. \end{proof}

\section{Computational Completeness}

\begin{theorem}\label{computational completeness} For any propositional modal formula $\phi_0$,
\begin{enumerate}
\item If $w\nVdash\phi_0$ for some world $w$ of a Kripke model $K$, then $\{\xi_u\}_{u\in U}\nvDash\phi_0$ for any enumeration $\{\xi_u\}_{u\in U}$ of nondeterministic partial recursive functions. 
\item If $w\nVdash\phi_0$ for some world $w$ of a deterministic Kripke model $K$, then $\{\xi_u\}_{u\in U}\nvDash\phi_0$ for any enumeration $\{\xi_u\}_{u\in U}$ of deterministic partial recursive functions.
\end{enumerate} 
\end{theorem}
\begin{proof} 
The two parts of this theorem will be proven simulteniously. Suppose $w_1\nVdash\phi_0$ for some world $w_1$ of the Kripke model $K$. Let $\{w_1,\dots,w_n\}$ be all worlds of this  Kripke model. Consider functions $f_i(x_1,\dots,x_n)$ such that
$$\xi_{f_i(x_1,\dots,x_n)}(u) =\{x_k\;|\; \exists j\;(u=x_j \wedge w_j\rightarrow_{w_i}w_k)\}.$$
Note that if Kripke model $K$ is deterministic, then $w_k$, mentioned in the above definition, is unique. Thus, partial recursive function $\xi_{f_i(x_1,\dots,x_n)}$ is deterministic. No matter if model $K$ is deterministic or nondeterministic, let us consider fixed points $u_1,\dots,u_n$ of functions $f_1,\dots,f_n$ whose existence follows from Theorem~\ref{Kleene}. Also, let valuation $*$ be defined on propositional variables as follows: $*(p)=\{u_i\;|\; w_i\Vdash p\}$.

\begin{lemma}\label{computational main} For any propositional modal formula $\phi$ and any $1\le i\le n$,
$$u_i \in \phi^* \;\;\;\;\; \Longleftrightarrow \;\;\;\;\; w_i \Vdash \phi.$$
\end{lemma}
\begin{proof} Induction on the complexity of formula $\phi$. By the definition of $*$, the  lemma is true for propositional variables. We will consider the only non-trivial inductive case: $\phi=\phi_1\rhd\phi_2$.
\begin{enumerate}

\item[$\Rightarrow$] Suppose $w_i \nVdash \phi_1\rhd\phi_2$. Thus, by Definition~\ref{forcing}, there are $j$ and $k$ such that $w_j\Vdash\phi_1$, $w_j\rightarrow_{w_i}w_k$, and for any $k'$ such that $w_j\rightarrow_{w_i}w_{k'}$, we have $w_{k'}\nVdash\phi_2$ . Thus, $u_k\in\xi_{f_i(u_1,\dots,u_n)}(u_j)$ and, at the same time, $w_{k'}\nVdash\phi_2$ for any $k'$ such that $u_{k'}\in \xi_{f_i(u_1,\dots,u_n)}(u_j)$. Hence $\xi_{f_i(u_1,\dots,u_n)}(u_j)$ is not empty and, by the induction hypothesis, $$\xi_{f_i(u_1,\dots,u_n)}(u_j)\cap\phi_2^*=\varnothing$$
By the choice of elements $u_1,\dots,u_n$, they are fixed points of functions $f_1,\dots,f_n$. Hence, $\xi_{u_i}(u_j)$ is not empty and $\xi_{u_i}(u_j)\cap\phi_2^*=\varnothing$. At the same time, by the induction hypothesis, $u_j\in\phi_1^*$. Thus, 
$$\neg\forall u\in\phi_1^*(\xi_{u_i}(u)\neq\varnothing\rightarrow
\xi_{u_i}(u)\cap\phi_2^*\neq\varnothing).$$
Therefore, by Definition~\ref{semantics}, $u_i\notin(\phi_1\rhd\phi_2)^*$.

\item[$\Leftarrow$] Assume that $u_i\notin(\phi_1\rhd\phi_2)^*$. Thus, by Definition~\ref{semantics}, there is an element $y\in U$ such that $y\in\phi_1^*$, $\xi_{u_i}(y)\neq\varnothing$, and $\xi_{u_i}(y)\cap\phi_2^*=\varnothing$. Note that since $\xi_{u_i}\equiv\xi_{f_i(u_1,\dots,u_n)}$, we can conclude that $\xi_{f_i(u_1,\dots,u_n)}(y)$ is also non-empty. This, by the definition of $f_i$ can happen only if $y=u_j$ for some $0\le j\le n$. In this case, by the same definition, $\xi_{u_i}(u_j)=\xi_{f_i(u_1,\dots,u_n)}(u_j)=\{u_k\;|\;w_j\rightarrow_{w_i}w_k\}$. Given that $y\in\phi_1^*$ and $\xi_{u_i}(y)\cap\phi_2^*=\varnothing$, we can conclude, by the induction hypothesis, that $w_j\Vdash\phi_1$ and $w_k\nVdash\phi_2$ for any $k$ such that $w_j\rightarrow_{w_i}w_k$. Therefore, by Definition~\ref{forcing}, $w_i\nVdash\phi_1\rhd\phi_2$.
\end{enumerate}
\end{proof}
To finish the proof of Theorem~\ref{computational completeness}, note that $w_1\nVdash\phi_0$ implies, by Lemma~\ref{computational main}, that $u_1\notin\phi_0^*$. Therefore, $\{\xi_u\}_{u\in U}\nvDash\phi_0$.  
\end{proof}

\begin{theorem}\label{grand}
For any propositional modal formula $\phi$ and any enumeration $\{\xi_u\}_{u\in U}$ of nondeterministic partial recursive functions, the following statements are equivalent:
\begin{enumerate}
\item $\{\xi_u\}_{u\in U}\vDash\phi$,
\item $w\Vdash\phi$ for every world $w$ of any Kripke model,
\item $\vdash_\Re\phi$.
\end{enumerate} 
\end{theorem}
\begin{proof}
Statement 1 implies statement 2 by Theorem~\ref{computational completeness}. Statement 2 implies statement 3 by  Theorem~\ref{Kripke completeness}. Statement 3 implies statement 1 by Theorem~\ref{soundness}. 
\end{proof}

\begin{corollary}
Modal logic $\Re$ is decidable.
\end{corollary}

\begin{theorem}\label{grand 2}
For any propositional modal formula $\phi$ and any enumeration $\{\xi_u\}_{u\in U}$ of deterministic partial recursive functions, the following statements are equivalent:
\begin{enumerate}
\item $\{\xi_u\}_{u\in U}\vDash\phi$,
\item $w\Vdash\phi$ for every world $w$ of any deterministic Kripke model,
\item $\vdash_{\Re_d}\phi$.
\end{enumerate} 
\end{theorem}
\begin{proof} The same as the proof of Theorem~\ref{grand}. 
\end{proof}

\begin{corollary}
Modal logic $\Re_d$ is decidable.
\end{corollary}

\section{Conclusions}

In this paper we have introduced two modal logics of partial recursive functions, gave their complete axiomatizations, and proved decidability of both logics. These results, of course, depend on the exact interpretation of connective $\rhd$ as given in Definition~\ref{semantics}. Let us consider two natural alternatives to this interpretation.

First of all, there are at least two different ways to define partial nondeterministic functions from set $A$ to set $B$. One approach is to require that all computational paths that start with an element in $A$ either do not terminate or terminate in $B$. The second approach is to say that if the terminating paths exist, then at least one of them ends in $B$. The second approach is normally used to define computation of a nondeterministic finite automaton and it is the approach adopted in Definition~\ref{semantics} of this paper. It is also possible to consider the logic of nondeterministic partial computable functions under the first approach. One can easily see that not only are all axioms of logic $\Re$ valid in this situation, but the axiom A4 of logic $\Re_d$ is valid too. Simple review of the given above completeness proof for logic $\Re_d$ shows that the same proof establishes completeness of $\Re_d$ as a logic of nondeterministic partial functions under the second approach.  

Secondly, one can define $(\phi\rhd\psi)^*$ to be the set of all {\em total} recursive functions from $\phi^*$ to $\psi^*$.  This definition seems to be especially appropriate given that under Curry-Howard isomorphism implication in the intuitionistic logic corresponds to the type of total recursive functions. In the case of modal logics of recursive functions, transition from partial to total functions is not trivial. Indeed, if $(\phi\rhd\psi)^*$ is interpreted as the set of all total (deterministic or nondeterministic) recursive functions from $\phi^*$ into $\psi^*$, then let's consider unary modality $\Diamond\phi\equiv\neg(\phi\rhd\bot)$. Note that a function from $\phi^*$  to $\varnothing$ exists only if $\phi^*$ is empty. Thus, set $(\Diamond\phi)^*$ is equal to the entire universe $U$ if set $\phi^*$ contains at least one element and set $(\Diamond\phi)^*$ is empty if $\phi^*$ is empty. The ability to define $\Diamond$ in the logics of total recursive functions makes it possible to express many properties that can not be expressed in the logics of partial functions. For example, formula
$\Diamond(\phi\rhd\psi)\wedge\Diamond(\psi\rhd\chi)\rightarrow\Diamond(\phi\rhd\psi)$ states, essentially, that the set of total functions is closed with respect to composition. A complete description of logics of total functions remains an open question.

\bibliography{naumov}

\begin{thebibliography}{16}
\expandafter\ifx\csname natexlab\endcsname\relax\def\natexlab#1{#1}\fi
\expandafter\ifx\csname url\endcsname\relax
  \def\url#1{{\tt #1}}\fi

\bibitem[Constable et~al.(1986)]{Constable1986}
R.~L. Constable et~al.
\newblock {\em Implementing {M}athematics with {N}uprl {P}roof {D}evelopment
  {S}ystem}.
\newblock Prentice Hall, 1986.

\bibitem[Coquand and Paulin(1990)]{MR91h:03012}
T.~Coquand and C.~Paulin.
\newblock Inductively defined types.
\newblock In {\em COLOG-88 (Tallinn, 1988)}, pages 50--66. Springer, Berlin,
  1990.

\bibitem[Curry(1934)]{Curry1934}
H.~B. Curry.
\newblock Functionality in combinatory logic.
\newblock {\em Proc. Nat. Acad. Sci. U. S. A.}, 20:\penalty0 584--590, 1934.

\bibitem[Curry(1942)]{MR3:289d}
H.~B. Curry.
\newblock The combinatory foundations of mathematical logic.
\newblock {\em J. Symbolic Logic}, 7:\penalty0 49--64, 1942.

\bibitem[Curry and Feys(1958)]{MR20:817}
H.~B. Curry and R.~Feys.
\newblock {\em Combinatory logic. {V}ol. {I}}.
\newblock North-Holland Publishing Co., Amsterdam, 1958.

\bibitem[Fairtlough and Mendler(1997)]{MR98j:03045}
M.~Fairtlough and M.~Mendler.
\newblock Propositional lax logic.
\newblock {\em Inform. and Comput.}, 137\penalty0 (1):\penalty0 1--33, 1997.
\newblock ISSN 0890-5401.

\bibitem[Harel et~al.(2000)Harel, Kozen, and Tiuryn]{MR2001i:68110}
D.~Harel, D.~Kozen, and J.~Tiuryn.
\newblock {\em Dynamic logic}.
\newblock MIT Press, Cambridge, MA, 2000.
\newblock ISBN 0-262-08289-6.

\bibitem[Howard(1980)]{MR82g:03094}
W.~A. Howard.
\newblock The formulae-as-types notion of construction.
\newblock In {\em To H. B. Curry: essays on combinatory logic, lambda calculus
  and formalism}, pages 480--490. Academic Press, London, 1980.

\bibitem[Kopylov and Nogin(2001)]{Kopylov2001}
A.~Kopylov and A.~Nogin.
\newblock Markov's principle for propositional type theory.
\newblock In L.~Fribourg, editor, {\em Computer Science Logic: 15th
  International Workshop, CSL 2001. 10th Annual Conference of the EACSL (Paris,
  France, 2001)}, volume 2142 of {\em Lecture Notes in Computer Science}, pages
  570--584. Springer, 2001.

\bibitem[McKinsey and Tarski(1944)]{MR5:211f}
J.~C.~C. McKinsey and A.~Tarski.
\newblock The algebra of topology.
\newblock {\em Ann. of Math. (2)}, 45:\penalty0 141--191, 1944.

\bibitem[Mendler(1991)]{MR92h:03020}
N.~P. Mendler.
\newblock Inductive types and type constraints in the second-order lambda
  calculus.
\newblock {\em Ann. Pure Appl. Logic}, 51\penalty0 (1-2):\penalty0 159--172,
  1991.
\newblock ISSN 0168-0072.
\newblock Second Annual IEEE Symposium on Logic in Computer Science (Ithaca,
  NY, 1987).

\bibitem[Moszkowski and Manna(1984)]{MR778950}
B.~Moszkowski and Z.~Manna.
\newblock Reasoning in interval temporal logic.
\newblock In {\em Logics of programs (Pittsburgh, Pa., 1983)}, volume 164 of
  {\em Lecture Notes in Comput. Sci.}, pages 371--382. Springer, Berlin, 1984.

\bibitem[Naumov(2003)]{Naumov03}
P.~Naumov.
\newblock An extension of the classical propositional logic by type
  constructors.
\newblock {\em The Bulletin of Symbolic Logic}, 9\penalty0 (2):\penalty0
  254--255, 2003.

\bibitem[Naumov(2004)]{Naumov04}
P.~Naumov.
\newblock Logic of subtyping.
\newblock {\em Theoretical Computer Science}, 2004.
\newblock (to appear).

\bibitem[Rogers(1987)]{MR88b:03059}
H.~Rogers, Jr.
\newblock {\em Theory of recursive functions and effective computability}.
\newblock MIT Press, Cambridge, MA, second edition, 1987.
\newblock ISBN 0-262-68052-1.

\bibitem[Smith(1995)]{Smith1995}
S.~Smith.
\newblock Hybrid partial-total type theory.
\newblock {\em Internat. J. Found. Comput. Sci.}, 6:\penalty0 235--263, 1995.
\newblock ISSN 0129-0541.

\end{thebibliography}

\end{document}